%% file: main.tex
\documentclass[lettersize,journal]{IEEEtran}

\usepackage{amsmath,amssymb}
\usepackage{enumitem}
\usepackage{xcolor}
\usepackage[T1]{fontenc}
\usepackage{graphicx}
\usepackage{booktabs}
\usepackage[table]{xcolor}
\usepackage{adjustbox}
\usepackage{arydshln}
\usepackage[hidelinks,hypertexnames=false]{hyperref}

\usepackage{amsmath,amsfonts}
\usepackage{algorithmic}
\usepackage{algorithm}
\usepackage{array}
\usepackage[caption=false,font=normalsize,labelfont=sf,textfont=sf]{subfig}
\newcommand{\centercaption}[1]{%
  \begingroup
  \makeatletter
  \long\def\@makecaption##1##2{%
    \vskip\abovecaptionskip
    \centerline{\normalfont\footnotesize ##1.~##2}%
    \vskip\belowcaptionskip}%
  \makeatother
  \caption{#1}%
  \endgroup
}
\usepackage{textcomp}
\usepackage{stfloats}
\usepackage{url}
\usepackage{verbatim}
\usepackage{graphicx}
\usepackage{cite}
\usepackage{xspace}
\newcommand{\obfscore}{\ensuremath{s}\xspace}

\begin{document}

\title{Obfuscation as an Effective Signal for Prioritizing Cross-Chain Smart Contract Audits: Large-Scale Measurement and Risk Profiling}

\author{Yao~Zhao,
Zhang~Sheng\textsuperscript{\dag},
Shengchen~Duan,
Shen~Wang,
Daoyuan~Wu,
Zhiyuan~Wan
\thanks{Yao Zhao is with the Department of Aeronautical and Aviation Engineering, The Hong Kong Polytechnic University, Hong Kong, China.}%
\thanks{Zhang Sheng is with The State Key Laboratory of Blockchain and Data Security, Zhejiang University, Hangzhou, China (e-mail: dcszhang@foxmail.com). \textsuperscript{\dag}Corresponding author.}%

\thanks{Shengchen Duan is with Singapore Management University, Singapore.}%
\thanks{Shen Wang is with Final Round AI.}%
\thanks{Daoyuan Wu is with Lingnan University, Hong Kong SAR, China.}%
\thanks{Zhiyuan Wan is with The State Key Laboratory of Blockchain and Data Security, Zhejiang University (e-mail: wanzhiyuan@zju.edu.cn).}%
}

\maketitle

\begin{abstract}
Obfuscation raises the interpretation cost of smart-contract auditing, yet its signals are hard to transfer across chains. We present \textsc{HObfNET}, a fast surrogate of \textsc{ObfProbe}, enabling million-scale cross-chain scoring. The model aligns with tool outputs on Ethereum (PCC 0.9158, MAPE 8.20\%) and achieves 8--9 ms per contract, yielding a 2.3k--5.2k$\times$ speedup. Across BSC, Polygon, and Avalanche, we observe systematic score drift, motivating within-chain percentile queues (p99 as the main queue, p99.9 as an emergency queue). The high-score tail is characterized by rare selectors, external-call enrichment, and low signature density, supporting secondary triage. Cross-chain reuse is tail-enriched and directionally biased from smaller to larger ecosystems. On two publicly alignable cross-chain spillover cases, both fall into the p99 queue, indicating real-world hit value. We deliver a two-tier audit queue and a cross-chain linkage workflow for practical security operations.
\end{abstract}

\begin{IEEEkeywords}
cross-chain obfuscation signals, smart-contract auditing, HObfNET, surrogate model, operationalizing security metrics, auditor triage, security operations workflow
\end{IEEEkeywords}
\input{Text/introduction}
\input{Text/background}
\input{Text/methodology}
\input{Text/results}
\input{Text/related_work}
\input{Text/discussion}
\input{Text/conclusion}

\bibliographystyle{IEEEtran}
\bibliography{references}

\end{document}

%% file: Text/introduction.tex
\section{Introduction}
Auditing smart contracts at scale is a triage problem: there are millions of contracts, but only limited analyst time. Obfuscation makes this harder by deliberately rewriting bytecode while keeping functionality the same, so transfer logic becomes difficult to inspect or explain. A recent Ethereum study by Sheng Z. et al.~\cite{sheng2025obfuscated} provides the first systematic characterization of obfuscated fund transfers. They show that obfuscation is routine but risk concentrates in the most heavily obfuscated contracts; within the heavily obfuscated Ethereum subset, they identify 463 high-risk cases including MEV bots, Ponzi schemes, fake decentralization, and extreme centralization, placing roughly \$100M at risk, and obfuscated scams exhibit higher peak losses than non-obfuscated ones. They also show that obfuscation can cripple existing detectors; for example, SourceP recall drops from 0.80 to 0.12, indicating that attackers can hide intent through obfuscation. Complementary work on closed-source contracts shows that obfuscation can conceal critical vulnerabilities rather than provide protection~\cite{yang2025obscurity}. At the same time, advances in bytecode analysis such as decompilation and function recovery enable large-scale measurement~\cite{grech2019gigahorse,he2023neuralfebi}. Together, these results establish obfuscation as a security-critical signal and point to a practical need: large-scale security operations must prioritize, and obfuscation scores are a natural ranking signal.

Now the EVM ecosystem is multi-chain. Contracts are routinely redeployed across BSC, Polygon, and Avalanche, and templates (including malicious ones) move across chains. This leads to a simple but urgent audit question: \emph{can Ethereum’s high-obfuscation cutoff be reused as a universal screening threshold for other chains?} If the cutoff fails to transfer, cross-chain audit queues will either overflow or miss high-risk candidates, leading to alert fatigue or blind spots. Code reuse is pervasive on Ethereum~\cite{he2020clones,chen2021codereuse}, multi-chain analyses reveal copy-and-paste deployments with non-trivial behavioral mismatches~\cite{wang2025equivguard}, and malicious templates such as sniper bots migrate across chains~\cite{cernera2025sniperbots}. Cross-chain tracing studies and industry reports further show that cross-chain fund migration and laundering are widespread~\cite{lin2025abctracer,trm2023illicit}. These observations imply that obfuscation signals may propagate across chains, yet their thresholds may not.

Even if we can build a main audit queue, “high score” alone is not enough—we must understand what the queue actually contains. If the tail is dominated by standard Ethereum Request for Comments (ERC)/NFT or proxy templates, the audit strategy should differ from one dominated by low-visibility, specialized logic. This calls for a clear structural profile of the high-score tail and simple, executable cues for secondary triage.

There is also a coordination problem. If high-score templates diffuse across chains, risk can spill over and requires cross-chain linkage. Cross-chain interoperability and fund migration complicate incident pathways: vulnerabilities may occur on one chain while funds move cross-chain for laundering or redistribution; meanwhile, the same template can be redeployed elsewhere. Therefore, obfuscation scores must be validated against real incidents rather than remain purely distributional signals.

The bottleneck is scale. Obfuscation tools are expensive to run. Based on our measurements, Sheng et al.'s ObfProbe can take weeks of runtime for a single chain and consume substantial compute~\cite{sheng2025obfuscated}, making cross-chain analysis slow and difficult to iterate. We therefore train a fast surrogate model, \textsc{HObfNET}, on Ethereum ObfProbe labels and use it to score large corpora on BSC, Polygon, and Avalanche.

\begin{figure*}[t]
  \centering
  \includegraphics[width=0.95\textwidth]{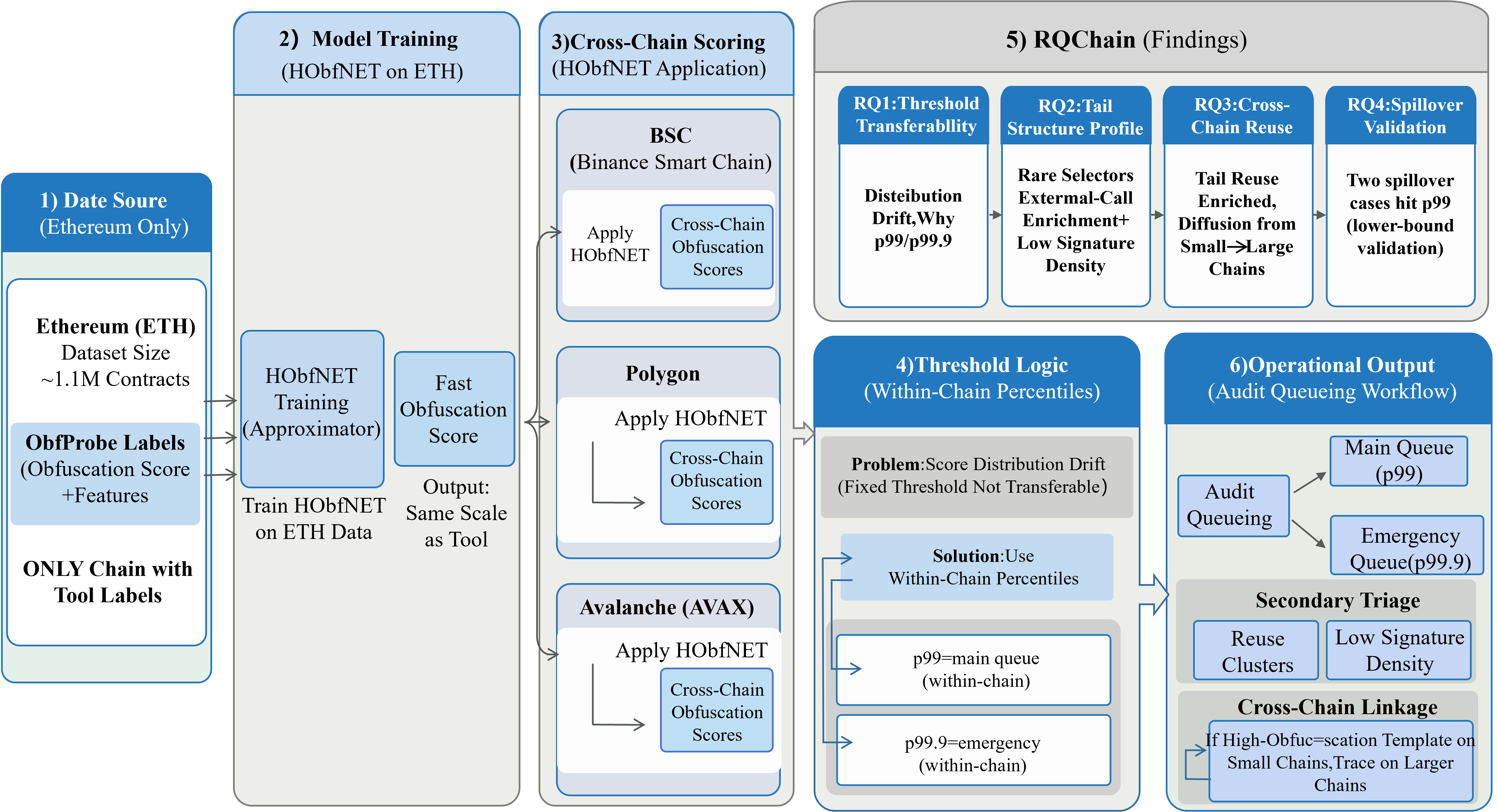}
  \caption{\textbf{End-to-end panorama.} Ethereum provides ObfProbe labels for training \textsc{HObfNET}; the model scores BSC/Polygon/Avalanche, enabling RQ1--RQ4 analysis and a two-tier audit queue with cross-chain linkage.}
  \label{fig:overview}
\end{figure*}

Figure~\ref{fig:overview} summarizes the end-to-end workflow: Ethereum provides tool labels for training; \textsc{HObfNET} produces cross-chain scores; RQ1--RQ4 explain thresholding, tail risk structure, reuse diffusion, and spillover validation; and the outputs form a practical audit queue with cross-chain linkage.

With cross-chain scores in hand, we focus on four practical questions:
\begin{enumerate}
  \item \textbf{RQ1 (Audit Thresholds):} Can Ethereum’s high-obfuscation cutoff be reused on other chains? If not, what within-chain cutoffs keep audit queues manageable?
  \item \textbf{RQ2 (Tail Profiling):} What is actually inside the high-score tail, and which simple cues help with secondary triage?
  \item \textbf{RQ3 (Cross-Chain Diffusion):} Do high-score templates reuse across chains, and does reuse flow from smaller to larger ecosystems?
  \item \textbf{RQ4 (Real-World Hits):} Do real cross-chain incidents fall into the high-score queue, validating its audit value?
\end{enumerate}

Our contributions are fivefold, in plain terms:
(1) We build a fast scoring model, \textsc{HObfNET}, that closely matches ObfProbe (MAPE 8.20\%, PCC 0.9158) but runs in 8--9 ms per contract (2.3k--5.2k$\times$ faster), so we can score millions of contracts.
(2) We show a single Ethereum cutoff does not carry over; instead we give per-chain cutoffs (p99/p99.9): ETH 18.07/22.69, BSC 16.82/19.74, Polygon 18.72/20.51, Avalanche 19.18/20.67. Using a fixed cutoff shifts queue sizes by 0.48\%--2.32\% (over- or under-including candidates).
(3) We open the “high-score queue” and explain what is inside: it is dominated by rare selectors, heavy external-call logic, and low signature density; on BSC the proxy indicator is enriched (0.25\%$\rightarrow$1.03\%). These are simple cues for secondary triage.
(4) We show that template copying across chains is real. When we only look at the high-score queue, the overlap between chains is much higher than in the full population (about 1.5--2$\times$), and the flow is mostly from smaller chains to larger ones. We can trace exact bytecode copies across chains.
(5) We validate with real incidents: the two publicly alignable cross-chain cases (Transit Swap DEX Hack, New Free DAO Flash Loan) both land in the top 1\% queue, showing the queue captures real-world risk.
We have open-sourced the model: \url{https://github.com/dcszhang/HObfNET}.

%% file: Text/background.tex
\section{Background and Preliminaries}
\label{sec:background}

\textbf{What is obfuscation in smart contracts?}
Smart contracts are deployed as EVM bytecode. Auditors and analysts often work with bytecode rather than source code. \emph{Obfuscation} deliberately makes program logic harder to understand by adding indirections, opaque predicates, junk code, and complex control-flow or dispatch patterns. These tricks distort the usual correspondence between bytecode structure and high-level intent, so even if the contract is functionally correct, its logic becomes expensive to reason about. In practice, obfuscation increases the \,``explanation cost'' of auditing and makes automated decompilation or rule-based inspection less reliable. This is why obfuscation is treated as a security-relevant signal rather than a purely cosmetic change~\cite{zhang2020obfuscation,yu2022bosc,yang2025obscurity}.

\textbf{Why obfuscation matters to security.}
Obfuscation can hide control-flow that would otherwise be straightforward to audit, and it can conceal malicious or risk-prone behavior behind opaque dispatch and indirect calls. Even when the code is not malicious, heavy obfuscation reduces transparency and raises review cost, which is a real operational risk when audit resources are limited. Recent studies also show that obscurity can coexist with vulnerabilities and complicate their detection, reinforcing the need to treat obfuscation as a risk-oriented signal rather than a purely stylistic artifact~\cite{yang2025obscurity}.

\textbf{Why a score?}
Obfuscation is not binary; it varies by degree. ObfProbe provides a scalar obfuscation score and a set of interpretable features derived from bytecode structure~\cite{sheng2025obfuscated}. Prior work shows that tool-based scoring can be accurate but slow at scale. Our goal is to preserve the meaning of the tool score while enabling million-scale cross-chain measurement.

\textbf{Why bytecode-level cues?}
Many deployed contracts do not publish verified source code, and bytecode-level analysis is often the only consistent view across chains. This motivates feature definitions that are observable directly from runtime bytecode. In our study, we rely on selectors, opcode patterns, and proxy-forwarding signatures because they are robust to missing ABI/source and are reproducible across chains.

\textbf{Why cross-chain?}
EVM-compatible chains share bytecode semantics, so the same scoring logic can be applied across chains. However, distributions can drift across ecosystems, making a single fixed cutoff unreliable. We therefore use within-chain percentiles to define practical audit queues (p99 for the main queue, p99.9 for an emergency queue), and then ask how these queues behave across chains~\cite{crosschain2020}.

\textbf{Key concepts used in later sections.}
We extract 4-byte function selectors from \texttt{PUSH4} immediates and define \emph{signature density} as the number of unique selectors per KB of bytecode. Low signature density typically indicates sparse interfaces or hidden logic. We also use a broad \emph{proxy indicator} to flag delegatecall-based forwarding patterns; this is a heuristic, not a strict EIP-1167 detector. These cues are used to build the tail structure profile and to support audit triage.

\textbf{Operational reading of the signals.}
Rare selectors and low signature density indicate low ABI visibility and higher explanation cost. Enrichment of external-call and return-data handling opcodes points to complex inter-contract orchestration. When these cues co-occur in the high-score tail, we treat them as a \emph{risk-oriented profile}: contracts that are harder to interpret and more likely to hide complex or risky behavior, which should be prioritized in audit queues.

\begin{center}
\fcolorbox{gray}{gray!10}{%
\parbox{0.95\linewidth}{\textbf{How this connects to the RQs.}
RQ1 examines whether Ethereum cutoffs transfer to other chains. RQ2 turns the high-score tail into a risk-oriented structural profile. RQ3 tests whether high-score templates diffuse across chains. RQ4 checks whether the queue hits real cross-chain spillover cases. Together, they form an actionable cross-chain audit workflow.}}
\end{center}

%% file: Text/methodology.tex
\section{Methodology and Model Evaluation}

\subsection{Task Definition}
Let $\mathcal{B} = \{\text{Ethereum}, \text{BSC}, \text{Polygon}, \text{Avalanche}\}$ be the chain set. For each chain $b \in \mathcal{B}$, we denote its contract corpus as $\mathcal{C}_b$. Each contract $c \in \mathcal{C}_b$ is represented by a bytecode sequence $x_c$, and our goal is to predict an obfuscation score $s_c \in \mathbb{R}$. We set $s_c = s_c^{\text{tool}}$, where $s_c^{\text{tool}}$ is produced by ObfProbe.

Our supervision signal comes from Sheng et al.'s ObfProbe on Ethereum. The tool outputs a scalar score $s_c^{\text{tool}}$ and a feature vector $F_c \in \mathbb{R}^K$ (transfer-path obfuscation features). We use the tool-defined obfuscation score as the learning target.

\subsection{Training Data and Preprocessing}
We use Ethereum contracts for training and validation because only Ethereum has tool-derived labels. Contracts from BSC, Polygon, and Avalanche are used only for cross-chain inference in Section~\ref{sec:cross-chain}.

\textbf{Bytecode normalization.} We strip the \texttt{0x} prefix and remove compiler metadata and constructor artifacts to obtain a canonical bytecode sequence.

\textbf{Segmentation.} The bytecode is mapped to integer tokens (0--255) and sliced into $N$ fixed-length segments of size $L$ with padding token 0. We maintain a validity mask $M \in \{0,1\}^N$ to distinguish real segments from padding. We report $L$ and $N$ in Section~\ref{sec:exp-setup}.

\textbf{Leakage control.} Within Ethereum, we perform train/validation/test splits at the bytecode-family level (near-duplicate clusters) to reduce clone leakage. Families are built using canonicalized bytecode fingerprints and clustering with a fixed similarity threshold to keep the split reproducible. We use a 7:2:1 split at the family level.

\subsection{Supervision Signals}
ObfProbe provides $K$ feature signals and a Z-score based obfuscation measure. We treat $s_c^{\text{tool}}$ (Z-score) as the primary regression target and reconstruct $F_c$ as an auxiliary objective. The Z-score is defined as:
\begin{equation}
  s_c^{\text{tool}} = \sum_{k=1}^{K} \frac{F_{c,k} - \mu_k}{\sigma_k},
\end{equation}
where $(\mu_k, \sigma_k)$ are computed on the Ethereum training set to avoid information leakage.

\subsection{Model Architecture}
\subsubsection{Overview}
\begin{figure*}[t]
  \centering
  \centerline{\includegraphics[width=0.9\textwidth]{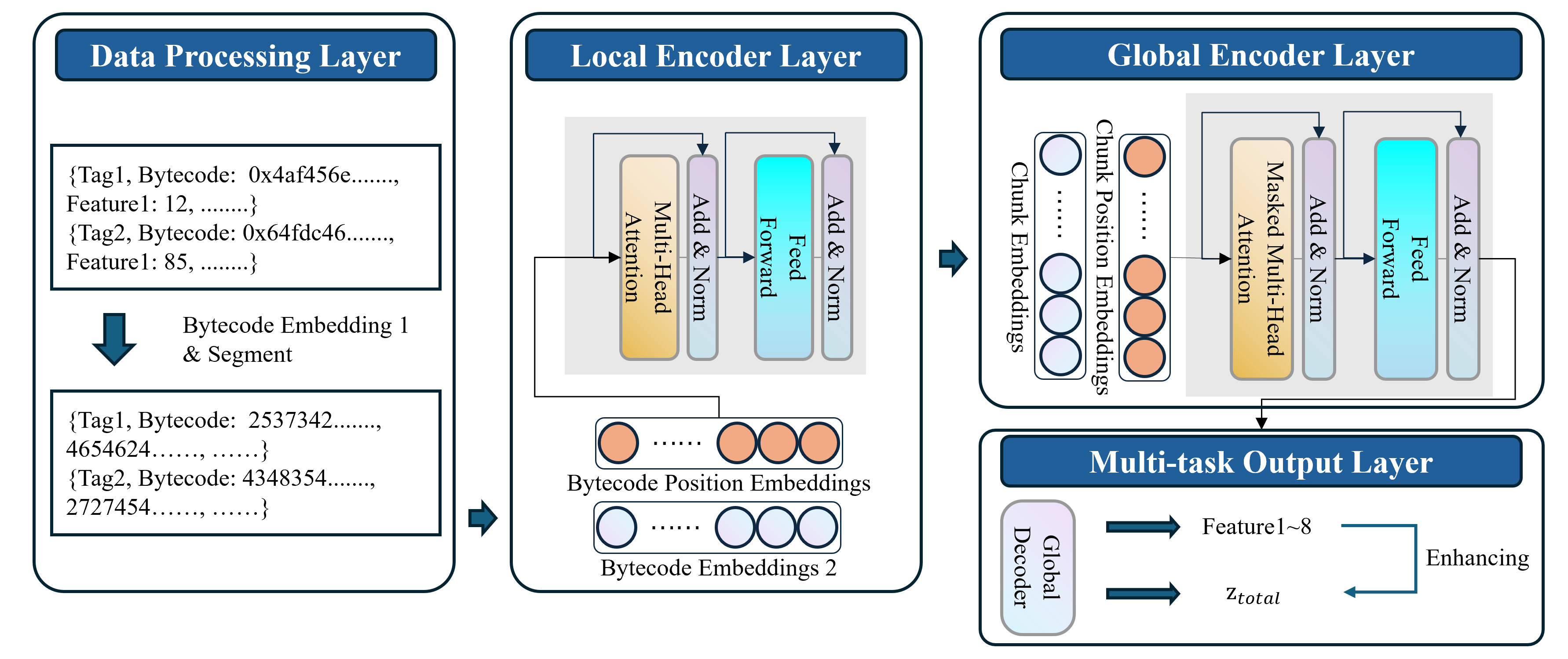}}
  \centercaption{The overall architecture of HObfNET: data processing, local encoder, global encoder, and a multi-task output layer that reconstructs domain features.}
  \label{Network}
\end{figure*}

Figure~\ref{Network} illustrates our model \textsc{HObfNET}, a hierarchical attention architecture designed for obfuscation scoring. The model follows a divide-and-conquer principle to handle extreme bytecode length and to capture both local opcode patterns and global contract structure. We report the number of layers and attention heads in Section~\ref{sec:exp-setup}.

\subsubsection{Local Encoder (local patterns)}
We embed each segment $x_i \in \mathbb{R}^L$ into a 256-dimensional space with positional embeddings to preserve opcode order. The local encoder captures short-range obfuscation artifacts (e.g., instruction-level patterns):
\begin{equation}
  H_{local}^{(i)} = \text{Transformer}_{enc}(\text{Embed}(x_{i}) + P_{local}).
\end{equation}

\subsubsection{Global Encoder (long-range dependencies)}
The global encoder models cross-segment dependencies that reflect contract-level obfuscation strategies. We apply multi-head attention with chunk-level positional encodings $P_{chunk}$ and allow bidirectional visibility to get the long-range obfuscation artifact $H_{global}^{(i)}$.

\subsubsection{Masked Mean Pooling (padding control)}
To avoid padding artifacts, we aggregate only valid segments using a masked mean:
\begin{equation}
  v_{contract} = \frac{\sum_{i=1}^{N} (H_{global}^{(i)} \cdot M_i)}{\sum_{i=1}^{N} M_i + \epsilon}.
\end{equation}
This prevents length imbalance from biasing the contract representation.

\subsubsection{Multi-task Enhancement (noise control and interpretability)}
We reconstruct domain features to regularize the latent space:
\begin{equation}
  \hat{F} = \text{MLP}_{rec}(v_{contract}).
\end{equation}
We compute an auxiliary score $\hat{s}^{\text{tool}}$ by applying the tool's Z-score formula to $\hat{F}$:
\begin{equation}
  \hat{s}^{\text{tool}} = \sum_{k=1}^{K} \frac{\hat{F}_k - \mu_k}{\sigma_k}.
\end{equation}
The final prediction fuses the deep representation, reconstructed features, and the auxiliary score:
\begin{equation}
  \hat{s} = \text{MLP}_{head}([v_{contract} \oplus \hat{F} \oplus \hat{s}^{\text{tool}}]).
\end{equation}

\subsection{Training Objective and Optimization}
We train end-to-end with a joint loss:
\begin{align}
  \mathcal{L}_{total} &= \lambda_{s} \mathcal{L}_{MSE}(s_c^{\text{tool}}, \hat{s}) + \lambda_{aux} \mathcal{L}_{MSE}(s_c^{\text{tool}}, \hat{s}^{\text{tool}}) \\
  &\quad + \lambda_{feature} \sum_{k=1}^{K} \mathcal{L}_{MSE}(F_{gt}^{(k)}, \hat{F}_k),
\end{align}
with $\lambda_{s}=1$, $\lambda_{aux}=0.1$, and $\lambda_{feature}=0.01$.

\subsection{Experimental Setup}
\label{sec:exp-setup}
We evaluate reliability from three angles: accuracy (mean absolute percentage error (MAPE), mean absolute error (MAE), mean squared error (MSE), and Pearson correlation coefficient (PCC) on the held-out Ethereum set), robustness (ablation and error analysis), and efficiency (throughput on GPU). We compare the full model with controlled variants (Standard Transformer, Gated Recurrent Unit (GRU)-based Hierarchical GRU Network (HGRU), and \textsc{HObfNET} without multi-task) to validate architectural choices. The training corpus follows the prior Ethereum dataset (~1.04M contracts, 2022-06 to 2024-08) with a 7:2:1 train/validation/test split at the family level. We set $L=512$, $N=32$, vocab size $=257$, $D_{\text{MODEL}}=256$, $\text{NUM\_LAYERS}=2$, $\text{N\_HEAD}=4$, the Global Encoder layer is set to $D_{\text{MODEL}}=256$, $\text{NUM\_LAYERS}=2$, $\text{N\_HEAD}=4$, and dropout $=0.1$. The $N\_HEAD$ is the number of heads in the multiheadattention models, and the $\text{NUM\_LAYERS}$ is the number of sub-encoder-layers. Training uses AdamW (lr $5 \times 10^{-4}$, weight decay $10^{-4}$), batch size $=24$, 20 epochs, and gradient clipping at 0.5 on an NVIDIA A100~\cite{zhang2025gaussian,sheng2025acoustic}. Inference uses batch size $=200$ on GPU. Incident alignment is address- and transaction-based (no explicit time window), and the sources are reported in RQ4.

\subsection{Evaluation Metrics}
To evaluate predictive accuracy and correlation, we report mean absolute percentage error (MAPE), mean absolute error (MAE), mean squared error (MSE), and Pearson correlation coefficient (PCC):
\begin{equation}
\begin{aligned}
    \mathcal{L}_{MAPE} &= \frac{1}{N} \sum_{i=1}^{N} \left| \frac{y_i - \hat{y}_i}{y_i} \right| \times 100\%, \\
    \mathcal{L}_{MAE} &= \frac{1}{N} \sum_{i=1}^{N} |y_i - \hat{y}_i|, \\
    \mathcal{L}_{MSE} &= \frac{1}{N} \sum_{i=1}^{N} (y_i - \hat{y}_i)^2, \\
    \mathcal{L}_{PCC} &= \frac{\sum_{i=1}^{N} (y_i - \bar{y})(\hat{y}_i - \bar{\hat{y}})}{\sqrt{\sum_{i=1}^{N} (y_i - \bar{y})^2} \sqrt{\sum_{i=1}^{N} (\hat{y}_i - \bar{\hat{y}})^2}},
\end{aligned}
\end{equation}
where $N$ is the number of test samples, $y_i$ and $\hat{y}_i$ are the tool score and prediction for sample $i$, and $\bar{y}, \bar{\hat{y}}$ are their means.

\subsection{Ablation Study}
To explicitly validate the effectiveness of the proposed hierarchical architecture and the domain-guided multi-task mechanism, we conducted an ablation study by comparing the full \textsc{HObfNET} model with three variants. The configuration of these variants is described as follows:
\begin{itemize}
    \item Standard Transformer:
    This variant removes the ``divide-and-conquer'' strategy. Instead of processing bytecode in chunks, it treats the entire smart contract as a single, flat sequence of bytes fed directly into a standard Transformer.
    \item Hierarchical GRU Network (HGRU) without Multi-Task:
    The Gated Recurrent Unit (GRU) is employed instead of the attention block to extract and expand the features.
    \item \textsc{HObfNET} without Multi-Task:
    This variant retains the hierarchical encoder structure (Local + Global Encoder) but removes the Multi-task module.
\end{itemize}

We report MAPE/MAE/MSE/PCC for \textsc{HObfNET} in Table~\ref{tab:ablation}.
First, the standard Transformer exhibits significantly lower performance compared to hierarchical models. This performance drop can be attributed to the forced truncation, which leads to the loss of critical logic located at the end of large contracts. It confirms that the hierarchical approach is essential for handling the extreme length variations of bytecode without information loss.
Second, HGRU underperforms the Transformer-based global encoder, indicating that attention captures long-range dependencies more effectively than recurrent aggregation.
The performance of \textsc{HObfNET} without the multi-task module indicates that removing auxiliary supervision leads to a noticeable increase in error rates. This suggests that latent features derived solely from raw bytecode are insufficient to fully capture high-level obfuscation patterns. Feature reconstruction and tool-consistent auxiliary scoring provide regularization, aligning representations with the tool-defined obfuscation metric.

\begin{table}[h]
    \centering
    \caption{\textbf{Ablation Study Results.} Comparison between the full model and its variants. Lower MAPE/MAE/MSE and higher PCC indicate better performance.}
    \label{tab:ablation}
    \footnotesize
    \setlength{\tabcolsep}{3pt}
    \renewcommand{\arraystretch}{0.9}
    \begin{tabular}{l|cccc}
        \toprule
        \textbf{Model} &  \textbf{MAPE} ($\downarrow$) & \textbf{MAE} ($\downarrow$) & \textbf{MSE} ($\downarrow$) & \textbf{PCC} ($\uparrow$) \\
        \midrule
        \textbf{Standard Transformer}      & 16.29 & 0.9521 & 2.7147 & 0.8466 \\
        \textbf{HGRU}                      & 14.28 & 0.8794 & 2.4511 & 0.8484 \\
        \textbf{HObfNET (w/o multi-task)}  & 13.02 & 0.8359 & 2.3371 & 0.8619 \\
        \midrule
        \textbf{HObfNET (Full Model)}      & 8.20  & 0.6341 & 1.4477 & 0.9158 \\
        \bottomrule
    \end{tabular}
\end{table}

\subsection{Efficiency Analysis}
\label{sec:efficiency}
Beyond accuracy, computational efficiency is critical for large-scale smart contract analysis. As a baseline, ObfProbe's Static Security Analysis (SSA) pipeline on Ethereum reports 93.55\% first-pass success under a 20s timeout; the first pass averages 19.66s (median 19.9s)~\cite{sheng2025obfuscated}. Failed contracts are rerun without a timeout cap, where the average runtime is about 45s (median 41s) and large/complex bytecodes often fall in the 20--80s range. Training \textsc{HObfNET} is performed on Ethereum; each epoch takes about 6.13 hours, with an additional 0.52 hours for validation on the same A100 server. For inference, we score BSC, Polygon, and Avalanche at scale.

\begin{table}[h]
    \centering
    \caption{\textbf{Efficiency Analysis.} ObfProbe vs.\ \textsc{HObfNET} inference time per contract (ms).}
    \label{tab:time_efficiency}
    \footnotesize
    \setlength{\tabcolsep}{4pt}
    \renewcommand{\arraystretch}{0.95}
    \begin{tabular}{l|r|c}
        \toprule
        \textbf{System / Chain} & \textbf{Contracts} & \textbf{Avg. time / contract (ms)} \\
        \midrule
        \textbf{ObfProbe (ETH, SSA)} & 1042923 & 19660 \\
        \textbf{HObfNET (BSC)}        & 2308899 & 8.67 \\
        \textbf{HObfNET (Polygon)}    & 288611 & 8.68 \\
        \textbf{HObfNET (Avalanche)}  & 96173 & 8.69 \\
        \bottomrule
    \end{tabular}
\end{table}

As shown in Table~\ref{tab:time_efficiency}, per-contract latency is stable (about 8--9 ms) across chains, indicating that the hierarchical encoder adds minimal overhead beyond bytecode normalization and chunked batching. For reference, ObfProbe reruns without a timeout cap average 45s (median 41s) and heavy bytecodes often fall in the 20--80s range. Compared with ObfProbe's Static Security Analysis (SSA) runtime on Ethereum, our per-contract inference is approximately 2.3k--5.2k$\times$ faster.

\subsection{Error Analysis}
We analyze prediction errors on the Ethereum validation set to understand when the surrogate deviates from the tool. Overall, the model attains MAPE 8.20\%, MAE 0.6341, MSE 1.4477, and PCC 0.9158, consistent with Table~\ref{tab:ablation}. The absolute error distribution is heavy-tailed: median 0.3008, p90 1.3340, p95 2.1702, and p99 5.4801.

\textbf{Length sensitivity.} We bin contracts by bytecode length quartiles and observe monotonic degradation as length grows (Table~\ref{tab:length_bins}): MAE/MAPE are 0.3747/8.18\% (0--4{,}802 bytes), 0.4829/9.52\% (4{,}802--11{,}244), 0.5952/10.97\% (11{,}244--19{,}186), and 1.0836/14.97\% (19{,}186--32{,}768).

\begin{table}[h]
    \centering
    \caption{\textbf{Length-binned errors are presented,} showing quartile bins of bytecode length and corresponding MAE/MAPE.}
    \label{tab:length_bins}
    \footnotesize
    \setlength{\tabcolsep}{4pt}
    \begin{tabular}{l|c|c}
        \toprule
        \textbf{Length bin (bytes)} & \textbf{MAE} & \textbf{MAPE} \\
        \midrule
        0--4{,}802 & 0.3747 & 8.18\% \\
        4{,}802--11{,}244 & 0.4829 & 9.52\% \\
        11{,}244--19{,}186 & 0.5952 & 10.97\% \\
        19{,}186--32{,}768 & 1.0836 & 14.97\% \\
        \bottomrule
    \end{tabular}
\end{table}

\textbf{Tail robustness.} We use two tail levels: p99 ($\tau_{eth}^{1\%}=18.07$, about 10{,}430 contracts) and an extreme tail p99.9 ($\tau_{eth}^{0.1\%}=22.69$, about 1{,}043 contracts). The reported errors are for the extreme tail (p99.9): MAE 2.0045 and MAPE 9.84\%, indicating that performance degrades but remains stable enough for tail-focused analysis.

\begin{center}
\fcolorbox{gray}{gray!10}{%
\parbox{0.95\linewidth}{\textbf{Plain summary.} This chapter builds \textsc{HObfNET}, a fast surrogate of ObfProbe for bytecode obfuscation scoring, shows it works through ablations and error analysis, and demonstrates that it is thousands of times faster than the tool baseline. These results justify using the model for large-scale cross-chain measurements in the next chapter.}}
\end{center}

%% file: Text/results.tex
\section{Results}
\label{sec:cross-chain}

This chapter uses the \textsc{HObfNET} model trained in Section~\ref{sec:exp-setup} to score deduplicated runtime bytecode on BSC, Polygon, and Avalanche. We obtain cross-chain obfuscation scores $\obfscore$ and answer RQ1--RQ4. For non-Ethereum chains, $\obfscore$ comes from model predictions; since the training objective aligns with the tool score, we treat them as comparable signals for distribution, tail, and reuse analyses.

\subsection{Data and Measurement Setup}
\label{subsec:setup-crosschain}

\textbf{Scale and time span.}
We normalize and deduplicate runtime bytecode (canonicalized runtime hash) and obtain 2,308,899 contracts on BSC, 288,611 on Polygon, and 96,173 on Avalanche. The coverage window is 2022-05 to 2024-10, referring to the collection/indexing period when we crawled and organized contract addresses and runtime bytecode from explorers/APIs.

\textbf{Score definition and tool features.}
The Ethereum training/validation file provides $\obfscore$ and tool-side features (feature1--feature7). We confirm that $\obfscore$ is the sum of these (standardized) feature terms; thus we treat it as the \emph{tool-aggregated scalar score} and use it as the aligned target across chains.

\textbf{Cross-chain sanity check.}
We perform a lightweight cross-chain spot-check by sampling 200k contracts from BSC and Polygon (and using all Avalanche contracts) and computing correlations between $\obfscore$ and bytecode length / signature counts. Correlations are consistently positive (0.46--0.72 for length; 0.10--0.27 for signature counts), indicating that the surrogate model preserves stable structural monotonicities across chains even without tool labels.

\textbf{Cross-chain tool alignment (10k per chain).}
To directly validate score comparability, we run ObfProbe on 10k contracts per chain and align tool scores with model predictions (Table~\ref{tab:tool_alignment}). The agreement is strong (PCC 0.96--0.97, MAE 0.45--0.47), supporting $\obfscore$ as a comparable cross-chain signal for distribution and tail analysis.

\begin{table}[!t]
  \centering
  \caption{\textbf{Cross-chain tool alignment (10k sample per chain).}}
  \label{tab:tool_alignment}
  \footnotesize
  \setlength{\tabcolsep}{5pt}
  \renewcommand{\arraystretch}{0.95}
  \begin{tabular}{l|r|c|c|c|c}
    \toprule
    \textbf{Chain} & \textbf{N} & \textbf{MAE} & \textbf{RMSE} & \textbf{MAPE} & \textbf{PCC} \\
    \midrule
    BSC & 10,000 & 0.45 & 0.93 & 5.7\% & 0.97 \\
    Polygon & 10,000 & 0.47 & 0.95 & 5.9\% & 0.96 \\
    Avalanche & 10,000 & 0.45 & 0.91 & 5.3\% & 0.97 \\
    \bottomrule
  \end{tabular}
\end{table}

\textbf{Feature definitions and heuristics.}
We extract 4-byte selectors from \texttt{PUSH4} immediates in runtime bytecode; the number of unique selectors is the \emph{signature count}. \emph{Signature density} is defined as $\#\text{unique selectors} / (\text{bytecode size in KB})$. ERC20/721 labels are assigned when bytecode contains a core selector set (ERC20: \texttt{totalSupply}, \texttt{balanceOf}, \texttt{transfer}, \texttt{approve}, \texttt{transferFrom}; ERC721: \texttt{balanceOf}, \texttt{ownerOf}, \texttt{approve}, \texttt{setApprovalForAll}, \texttt{transferFrom}, \texttt{safeTransferFrom}), used as a heuristic indicator rather than a definitive classification. The \emph{proxy indicator} marks contracts whose runtime bytecode exhibits delegatecall-based forwarding patterns (e.g., presence of \texttt{DELEGATECALL} with short forwarding dispatch), and is intentionally broader than strict EIP-1167 matching. For selector/opcode enrichment, we compute $\text{lift}(x)=p_{\text{tail}}(x)/p_{\text{all}}(x)$ with a small $\epsilon$ smoothing term in the denominator, and we only report patterns above a minimum occurrence threshold to avoid unstable lift ratios.

\textbf{Two-level tail definition.}
We adopt two tail levels: (i) \emph{main tail} as chain-level p99 (Top 1\%) for stable statistics; (ii) \emph{extreme tail} as p99.9 (Top 0.1\%) for robustness analysis. The corresponding score cutoffs are: main tail (ETH 18.07, BSC 16.82, Polygon 18.72, Avalanche 19.18) and extreme tail (ETH 22.69, BSC 19.74, Polygon 20.51, Avalanche 20.67).

\subsection{RQ1: Distribution Transferability and Threshold Drift}
\textbf{Question (starting point).} We begin with the most actionable audit question: \emph{are obfuscation-score distributions consistent across chains, and can Ethereum’s high-score cutoff be reused as a universal screening threshold?} This section provides a definitive answer and sets up why differences must be explained.

\subsubsection{Cross-chain distributions and quantiles}
Table~\ref{tab:score_quantiles} reports score quantiles. Medians are similar (around 5.0--5.4), but high quantiles diverge: BSC p90/p95/p99 = 9.60/12.45/16.82, notably lower than Polygon (13.50/15.66/18.72) and Avalanche (14.09/16.30/19.18). This indicates a lighter tail on BSC and heavier tails on Polygon/Avalanche.

\begin{table*}[!t]
  \centering
  \caption{\textbf{Score quantiles across chains (p99, p99.9).}}
  \label{tab:score_quantiles}
  \footnotesize
  \setlength{\tabcolsep}{4pt}
  \renewcommand{\arraystretch}{0.9}
  \begin{tabular}{l|r|ccccc}
    \toprule
    \textbf{Chain} & \textbf{N} & \textbf{p50} & \textbf{p90} & \textbf{p95} & \textbf{p99} & \textbf{p99.9} \\
    \midrule
    Ethereum & 1,042,923 & 5.03 & 9.37 & 12.44 & 18.07 & 22.69 \\
    BSC & 2,308,899 & 5.13 & 9.60 & 12.45 & 16.82 & 19.74 \\
    Polygon & 288,611 & 5.37 & 13.50 & 15.66 & 18.72 & 20.51 \\
    Avalanche & 96,173 & 5.21 & 14.09 & 16.30 & 19.18 & 20.67 \\
    \bottomrule
  \end{tabular}
\end{table*}

\subsubsection{What happens if we transfer the Ethereum cutoff (18.07)?}
We test direct transfer of the Ethereum cutoff 18.07 and compute $\Pr(\obfscore\ge 18.07)$ on other chains; Table~\ref{tab:tau_migration} reports the tail shares: BSC 0.48\%, Polygon 1.58\%, and Avalanche 2.32\%. Relative to Ethereum's 1.00\% baseline, BSC shrinks while Polygon/Avalanche inflate, showing that a fixed threshold does not translate consistently across chains.

From an operational perspective, this creates immediate audit distortion. With the ETH cutoff (18.07), BSC would flag only 11,079 contracts (0.48\%) instead of its chain-specific baseline of $\sim$23,089, \emph{missing about 12k high-obfuscation candidates}. In contrast, Polygon and Avalanche inflate to 4,553 (+58\%) and 2,229 (+132\%) candidates, respectively, dramatically increasing review load. Here, “high-obfuscation candidates” refers to the \emph{within-chain Top1\% (p99) candidate set}. Thus threshold drift is not a statistical curiosity—it translates directly into \emph{missed coverage} or \emph{alert overload}.

\textbf{Queue length and staffing cost.} Let the manual review throughput be $r$ contracts/day (or average review time $t$ minutes per contract). The candidate gap $\Delta n$ induced by threshold transfer directly converts to labor cost: $\Delta \text{days}=\Delta n / r$ or $\Delta \text{hours}=\Delta n \cdot t / 60$. For BSC, reusing the ETH cutoff reduces the queue by $\sim 1.2\times 10^4$ relative to the chain’s Top1\% baseline, i.e., $\approx 1.2\times 10^4/r$ days of review workload, but at the cost of excluding a comparable number of high-percentile candidates (potential misses).

\textbf{Actionable two-tier triage.} In practice, we recommend using $\obfscore$ as a triage signal rather than a single hard cutoff: a main queue for routine audits and continuous monitoring, and an emergency queue for the most extreme cases that warrant immediate reverse engineering and fund-flow tracing.
Accordingly, we recommend \emph{chain-specific, actionable} screening cutoffs: main screening scores of ETH 18.07, BSC 16.82, Polygon 18.72, and Avalanche 19.18 to maintain stable candidate volumes; and extreme-priority cutoffs of ETH 22.69, BSC 19.74, Polygon 20.51, and Avalanche 20.67 when prioritizing the most severe cases.

\textbf{Practical reminder.} Cross-chain spillovers are rare and may not always land in the extreme tail. We therefore treat p99 as the main queue and suggest maintaining a wider “watch band” (e.g., p95--p99) for continuous observation, while using the main/extreme cutoffs for resource-constrained prioritization.

Therefore, for cross-chain audit dashboards, it is often more practical to expose \emph{within-chain percentiles} as a unified scale: map each contract’s score to its position in the chain-specific distribution, and trigger alerts at fixed percentiles (e.g., p99/p99.9).

\begin{table*}[!t]
  \centering
\caption{\textbf{Absolute-threshold transfer (ETH cutoff = 18.07).}}
  \label{tab:tau_migration}
  \small
  \setlength{\tabcolsep}{6pt}
  \renewcommand{\arraystretch}{1.0}
  \begin{tabular}{l|r|c}
    \toprule
    \textbf{Chain} & \textbf{Tail count} & \textbf{Tail share} \\
    \midrule
    Ethereum & 10,430 & 1.00\% \\
    BSC & 11,079 & 0.48\% \\
    Polygon & 4,553 & 1.58\% \\
    Avalanche & 2,229 & 2.32\% \\
    \bottomrule
  \end{tabular}
\end{table*}

\subsubsection{Threshold Transfer Sensitivity Across Chain-Specific Cutoffs}
We compare cross-chain transfers of each chain’s main cutoff. Using the BSC cutoff (16.82) yields 3.27\% (Polygon) and 4.20\% (Avalanche), while using the Avalanche cutoff (19.18) yields only 0.17\% on BSC. Even among chain-specific main cutoffs, cross-chain transfer causes substantial shifts.

\subsubsection{Extreme Tail Robustness}
Transferring the Ethereum extreme cutoff (22.69) yields zero hits on all three chains, suggesting stronger chain-specificity in the extreme tail (or more conservative cross-chain scoring). Note that $\obfscore$ on non-Ethereum chains is model-predicted rather than tool-derived; thus ``zero hits'' may also reflect calibration/extrapolation uncertainty under cross-chain drift, further motivating within-chain percentiles for comparison. This does not mean other chains lack risky contracts; rather, absolute extreme thresholds do not transfer well. We therefore use the chain-specific main cutoffs as the primary tail definition and keep extreme cutoffs for robustness.

\medskip
\begin{center}
\fcolorbox{gray}{gray!10}{%
\parbox{0.95\linewidth}{\textbf{Finding 1 (RQ1).}
The Ethereum threshold cannot serve as a universal cross-chain screening line: differing chain distributions cause systematic queue shrinkage or inflation, leading to missed coverage or queue overload.
Audit practice should therefore rely on chain-specific percentile thresholds (main queue p99: 18.07/16.82/18.72/19.18; emergency queue p99.9: 22.69/19.74/20.51/20.67) or explicit calibration mappings, rather than a single fixed value.}}
\end{center}
This matters operationally: an over-wide threshold inflates review queues and audit budgets, while an over-strict threshold suppresses alerts and increases missed coverage. Based on our analysis, we provide chain-specific obfuscation audit score references for the four chains to support practical screening.

\subsection{RQ2: Tail Structure}
\textbf{Following RQ1.} If distributions differ, we ask \emph{what drives those differences}. Our goal is not just description: the tail profile should \emph{directly signal high risk}. In practice, the high-obfuscation tail is dominated by \emph{low-visibility, hard-to-audit logic} rather than standard ERC/NFTs or simple proxies. We therefore use ERC labels, a proxy indicator, and signature density to connect structure to risk: \emph{rare selectors + external-call enrichment + low signature density} indicate hidden logic, cross-contract orchestration, and high audit cost.

\subsubsection{ERC20/ERC721 presence in the tail}
Table~\ref{tab:erc_tail} compares overall vs. main-cutoff tail ERC shares. Overall ERC20 shares are 2.88\% (BSC), 0.35\% (Polygon), and 0.84\% (Avalanche), while overall ERC721 shares are 0.07\%, 3.42\%, and 0.55\%, respectively. In the tail, ERC20/ERC721 shares drop sharply: BSC ERC20 falls from 2.88\% to 0.17\% ($\sim$17$\times$), and Polygon/Avalanche ERC721 drops to $<0.01\%$ (order-of-magnitude decline).

\begin{table*}[!t]
  \centering
  \caption{\textbf{This figure illustrates ERC20/ERC721 shares, comparing overall distribution versus the main-cutoff tail.}}
  \label{tab:erc_tail}
  \footnotesize
  \setlength{\tabcolsep}{4pt}
  \renewcommand{\arraystretch}{0.9}
  \begin{tabular}{l|cc|cc}
    \toprule
    \textbf{Chain} & \textbf{ERC20\%} & \textbf{ERC721\%} & \textbf{Tail ERC20\%} & \textbf{Tail ERC721\%} \\
    \midrule
    BSC & 2.88 & 0.07 & 0.17 & 0.04 \\
    Polygon & 0.35 & 3.42 & $<0.01$ & $<0.01$ \\
    Avalanche & 0.84 & 0.55 & 0.21 & $<0.01$ \\
    \bottomrule
  \end{tabular}
\end{table*}

This indicates that the high-obfuscation tail is not dominated by standard ERC contracts. Given high sighash missingness and lower ABI visibility in obfuscated contracts, ERC labels are likely undercounted; we thus interpret this as ``tail contracts are less standard and less ABI-visible.'' From an audit perspective, focusing only on ERC asset contracts would systematically miss the high-obfuscation tail. The structural cues and exemplar checks below further support this interpretation.

\subsubsection{Proxy Indicator Prevalence}
\textbf{Scope note.} We distinguish two notions: (i) \emph{EIP-1167 minimal proxy} (strict template matching, extremely rare overall and in the tail, serving as a lower bound), and (ii) a broader \emph{proxy indicator} (heuristic patterns covering transparent/UUPS/custom forwarders, potentially over/under-counting). Table~\ref{tab:proxy_tail} reports (ii); strict minimal proxy is only a lower-bound reference and should not be interpreted as equivalent to EIP-1167 prevalence.
We mark contracts with a proxy indicator and report prevalence in overall and tail sets (Table~\ref{tab:proxy_tail}). The indicator is based on heuristic proxy bytecode patterns and forwarding behaviors, and is intentionally broad rather than a strict EIP-1167 detector; as such it may over/under-count. Proxy indicators remain rare overall (sub-1\%); under the main cutoff, BSC shows clear enrichment (0.25\%$\rightarrow$1.03\%) while Polygon/Avalanche are flat or slightly lower. This suggests proxy patterns are not the dominant tail component, but can be active in certain chains' high-score regions. Therefore, ERC/Proxy alone still cannot explain the tail, motivating structural cues below.

\begin{table}[!t]
  \centering
\caption{\textbf{Broader proxy indicator prevalence (overall vs. main-cutoff tail).}}
  \label{tab:proxy_tail}
  \small
  \setlength{\tabcolsep}{4pt}
  \resizebox{\linewidth}{!}{%
  \begin{tabular}{l|c|c}
    \toprule
    \textbf{Chain} & \textbf{Overall Proxy indicator\%} & \textbf{Tail Proxy indicator\%(main cutoff)} \\
    \midrule
    BSC & 0.25 & 1.03 \\
    Polygon & 0.69 & 0.52 \\
    Avalanche & 0.90 & 0.62 \\
    \bottomrule
  \end{tabular}
  }
\end{table}

\subsubsection{Selector/opcode enrichment: structural cues of the tail}
\textbf{Enrichment definition.} For each chain, we compute selector/opcode frequencies in the tail and overall sets, and define lift as $\text{lift}(x)=\frac{p_{\text{tail}}(x)}{p_{\text{all}}(x)}$. To avoid inflation from tiny counts, we only report patterns with counts above a minimum threshold (and apply $\epsilon$ smoothing to denominators); opcodes are counted by total occurrences.
We further compare selector and opcode distributions between the tail and the overall population, and extract the most enriched patterns (shown below). The tail is dominated by \emph{rare 4-byte selectors} with very high lift (10--50$\times$), rather than standard ERC interfaces, indicating non-standard or specialized logic. On the opcode side, tail contracts show clear enrichment of stack operations and external-call plumbing (e.g., DUP8--DUP11, RETURNDATASIZE/RETURNDATACOPY, GAS, STATICCALL), consistent with obfuscation and template-like logic.

\noindent\textbf{BSC:} selectors \texttt{0xfe9179cc} (45.0×), \texttt{0xd3e1c284} (34.2×), \texttt{0xfa483e72} (33.0×); opcodes: DUP10 (2.50×), RETURNDATASIZE (2.21×), GAS (2.19×).\\
\textbf{Polygon:} selectors \texttt{0xfce8337f} (52.6×), \texttt{0xef590ca5} (52.6×), \texttt{0xe6a43905} (32.6×); opcodes: STATICCALL (2.26×), RETURNDATACOPY (2.08×), GAS (2.04×).\\
\textbf{Avalanche:} selectors \texttt{0xdc7e0ce8} (15.2×), \texttt{0x37d20fff} (13.1×), \texttt{0x61d027b3} (9.9×); opcodes: DUP11 (1.97×), DUP10 (1.82×), STATICCALL (1.63×).

In addition, the most frequent tail selectors still include \texttt{0xf2fde38b} (transferOwnership) and \texttt{0x8da5cb5b} (owner), indicating that tail contracts mix governance/permission entry points with non-standard logic. Many high-lift selectors do not map to public standard interfaces, so we treat them as structural cues rather than functional labels.

\subsubsection{Manual inspection of top tail exemplars}
To provide illustrative exemplars, we manually inspect the highest-scoring contract from each chain. These contracts consistently exhibit \emph{very low signature density} (about 0.25/KB), consistent with the rare-selector and external-call enrichment signals, indicating low-visibility, template-like logic.

\textbf{Consistency with high-risk audit cues.}
The above signals match common “high explanation-cost” audit cues and point to risk: rare selectors and low signature density imply sparse, opaque interfaces; enriched external-call and return-data handling suggest complex inter-contract orchestration; and cross-chain template reuse enables fast propagation. In other words, the tail profile is not just structure—it is a \emph{high-risk signature}. We therefore treat the high-obfuscation tail as a \emph{higher-priority audit queue}.

\medskip
\begin{center}
\fcolorbox{gray}{gray!10}{%
\parbox{0.95\linewidth}{\textbf{Finding 2 (RQ2).}
The high-score tail is a \emph{high-risk signature}, not just a structural description. \emph{Rare selectors + external-call enrichment + low signature density} jointly indicate hidden logic, cross-contract manipulation, and high audit cost—exactly the traits we associate with risky contracts. These signals can be used directly to triage the RQ1 queue and prioritize likely fast-spreading high-risk templates.}}
\end{center}
\begin{center}
\fcolorbox{gray}{gray!10}{%
\parbox{0.95\linewidth}{\textbf{Audit playbook: secondary triage within the tail queue.}
Given the RQ1 chain-specific screening queue, prioritize contracts that:
(1) satisfy \emph{low signature density + high external-call enrichment + rare selectors};
(2) belong to cross-chain reuse clusters (same bytecode hash on multiple chains);
(3) expose \texttt{owner}/\texttt{transferOwnership} entry points with unusually sparse selector sets;
(4) hit the proxy indicator while implementation logic is opaque (trace the implementation).
This triage reduces alert fatigue and concentrates analyst time on templates with the highest explanation cost and propagation potential.}}
\end{center}

\subsection{RQ3: Cross-Chain Reuse and Propagation}
\textbf{Following RQ2.} RQ2 shows that the tail is dominated by a profile of “rare selectors + external-call enrichment + low signature density.” The key next question is whether these structural signals are driven by the \emph{same reusable templates propagating across chains}. If so, high-obfuscation risk exhibits cross-chain diffusion and calls for coordinated auditing. We test the existence, directionality, and amplification of cross-chain reuse in the high-obfuscation tail.

\subsubsection{Overall Reuse and Directional Asymmetry}
\begin{figure*}[!t]
  \centering
  \begin{minipage}{0.48\textwidth}
    \centering
    \includegraphics[width=\textwidth]{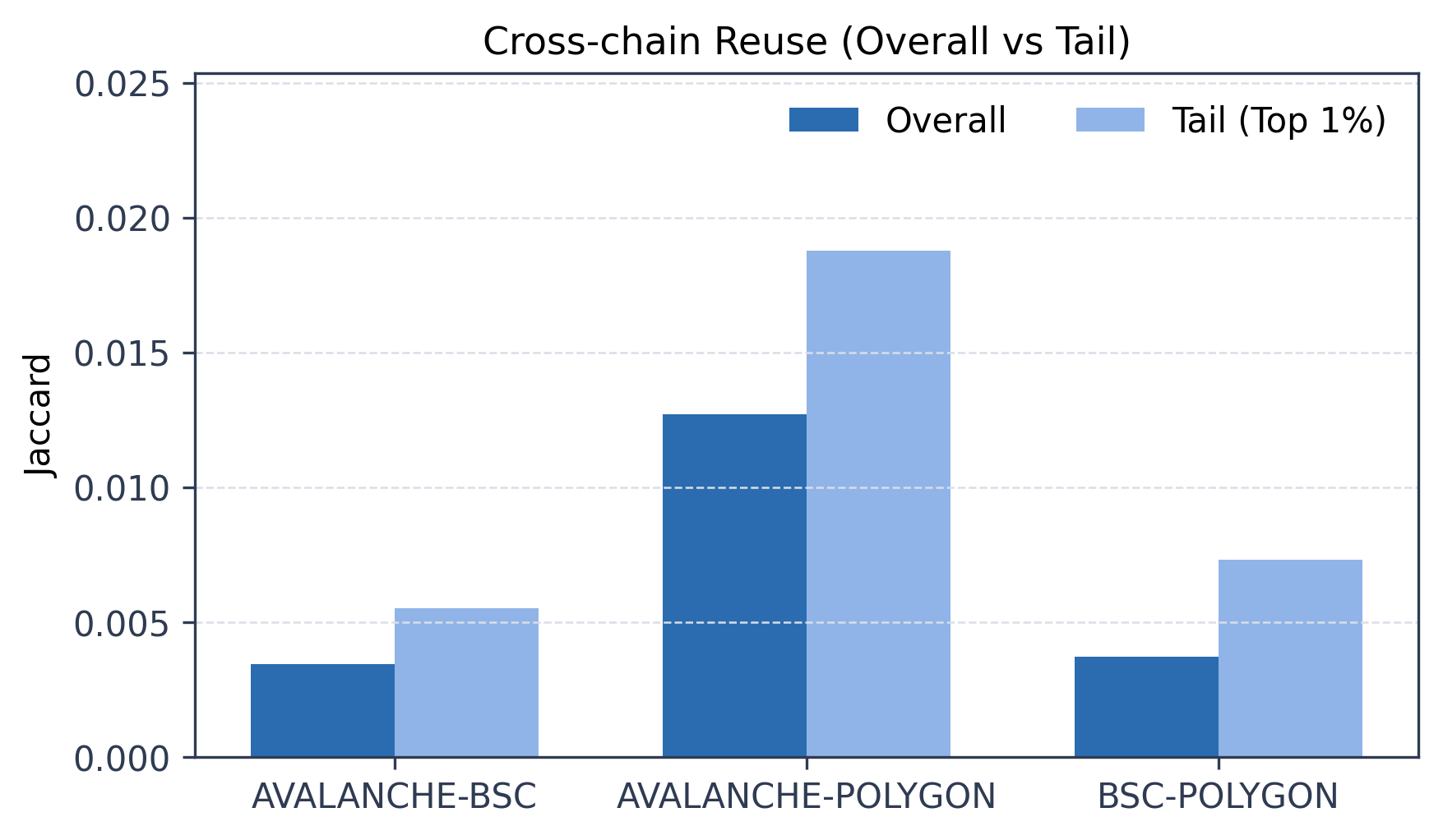}
  \end{minipage}\hfill
  \begin{minipage}{0.48\textwidth}
    \centering
    \includegraphics[width=\textwidth]{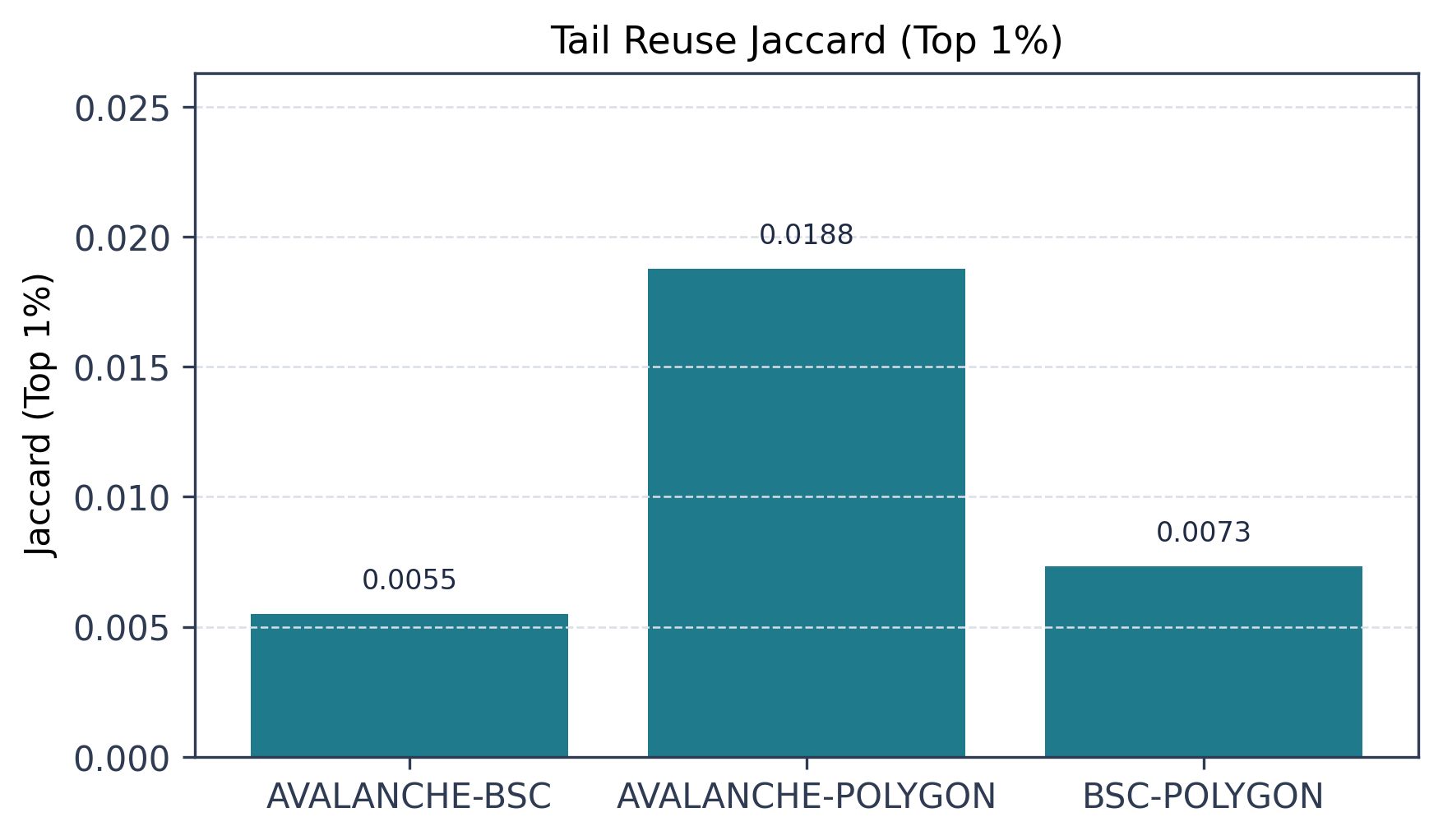}
  \end{minipage}
  \caption{Cross-chain reuse overlap (overall vs.\ tail Jaccard). Values are computed on deduplicated runtime bytecode hashes.}
  \label{fig:reuse_overlap}
\end{figure*}

\begin{figure}[!t]
  \centering
  \includegraphics[width=0.95\columnwidth]{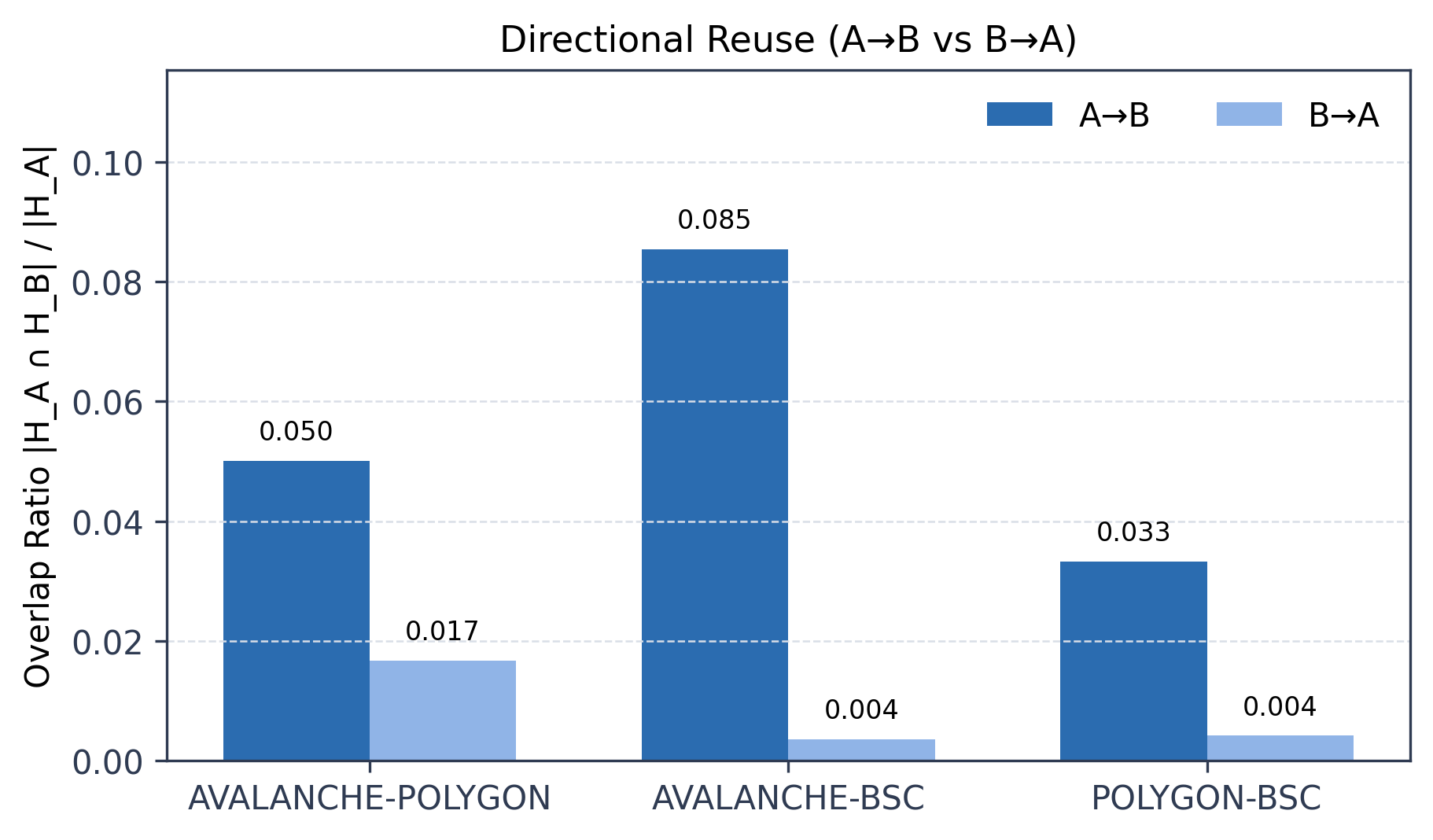}
  \caption{Directional overlap (small$\rightarrow$large higher), where $\text{overlap}(A\!\rightarrow\!B)=|H_A \cap H_B|/|H_A|$ and $H_A$ is the deduplicated hash set of chain $A$.}
  \label{fig:reuse_overlap_dir}
\end{figure}

We use deduplicated runtime bytecode hashes as reuse identifiers and measure cross-chain overlap via Jaccard and overlap coefficients (Table~\ref{tab:reuse_overlap}). Jaccard values are low (0.003--0.013), indicating that each chain has a largely distinct contract ecosystem. Figure~\ref{fig:reuse_overlap} contrasts overall vs.\ tail reuse; “tail Jaccard” is computed after extracting tail sets (Top1\%) within each chain and then computing hash Jaccard across tail sets, while “overall Jaccard” is computed on full deduplicated sets. Tail Jaccard is clearly higher than overall (roughly 0.0055--0.0188 vs 0.0034--0.0127), a 1.5--2.0$\times$ enrichment, with Polygon--Avalanche most pronounced; in some pairs it approaches a 2$\times$ increase (Fig.~\ref{fig:reuse_overlap}). Directionally, overlap is much higher from small$\rightarrow$large chains than the reverse (e.g., Avalanche$\rightarrow$BSC 8.55\% vs 0.36\%, Polygon$\rightarrow$BSC 3.33\% vs 0.42\%), where $\text{overlap}(A\!\rightarrow\!B)=\frac{|H_A \cap H_B|}{|H_A|}$ and $H_A$ is the deduplicated hash set of chain $A$ (Fig.~\ref{fig:reuse_overlap_dir}). This asymmetry suggests cross-chain reuse is closer to diffusion from smaller ecosystems to larger ones (e.g., copying templates into larger markets for greater visibility and interaction opportunities). Together with the RQ2 profile, this supports the mechanism that template-like, low-visibility contracts are more reusable and thus amplified in cross-chain tails.

\begin{table*}[!t]
  \centering
  \caption{\textbf{Overall cross-chain bytecode reuse analysis.}}
  \label{tab:reuse_overlap}
  \footnotesize
  \setlength{\tabcolsep}{4pt}
  \renewcommand{\arraystretch}{0.9}
  \begin{tabular}{l|c|c|c}
    \toprule
    \textbf{Pair} & \textbf{Jaccard} & \textbf{Small$\rightarrow$Large} & \textbf{Large$\rightarrow$Small} \\
    \midrule
    Avalanche--Polygon & 0.0127 & 5.01\% & 1.67\% \\
    Avalanche--BSC & 0.0034 & 8.55\% & 0.36\% \\
    Polygon--BSC & 0.0037 & 3.33\% & 0.42\% \\
    \bottomrule
  \end{tabular}
\end{table*}

\subsubsection{High-Obfuscation Template Reuse Across Chains}
We analyze reuse clusters (by bytecode hash) and select high-mean-$\obfscore$ clusters. These clusters tend to have long bytecode, low signature counts, and multi-chain coverage, aligning with the ``low ABI visibility'' profile from RQ2. The statistics show mean-$\obfscore$ shifting upward with greater chain coverage; both 2-chain and 3-chain clusters contain high-$\obfscore$ outliers (near $\sim$20), indicating strong cross-chain \emph{reusability/replicable deployment}: the same template reappears across chains with consistently high mean-$\obfscore$.

\textbf{Cross-chain reuse cases.}
We observe identical bytecode hashes reappearing across multiple chains, and some cases even show identical addresses across chains. These patterns align with template-driven deployment or deterministic address strategies (e.g., CREATE2), providing traceable evidence of cross-chain reuse. Identical-address reuse further suggests deterministic deployment or unified scripting; defensively, this enables \emph{cross-chain tracing and preemptive alerting} once a high-risk template is identified on any single chain.

\medskip
\begin{center}
\fcolorbox{gray}{gray!10}{%
\parbox{0.95\linewidth}{\textbf{Finding 3 (RQ3).}
Overall reuse is low, but tail reuse is markedly enriched (Jaccard up by $\sim$1.5--2.0$\times$) and directionally asymmetric (small$\rightarrow$large is far higher than the reverse). This suggests reuse is closer to diffusion from smaller ecosystems to larger ones, where high-score templates are copied into bigger markets and create cross-chain spillovers.}}
\end{center}

\begin{center}
\fcolorbox{gray}{gray!10}{%
\parbox{0.95\linewidth}{\textbf{Audit Action (RQ3).}
Once a high-score reuse cluster is detected on any chain, immediately search and trace the same bytecode hash across other chains to reduce lagged discovery costs.}}
\end{center}

\subsection{RQ4: Incident Alignment and Investigative Value}
\textbf{Following RQ3.} RQ4 asks one practical question: \emph{for the alignable cross-chain spillover cases, does the high-score queue hit real risk?} Public cross-chain spillovers are rare, so we use the two alignable cases as a \emph{lower-bound} validation. If they hit the queue, it indicates real-world value; if not, the queue is only a weak clue.

\subsubsection{Alignment overview}
We extract incident-related addresses from public reports and tx/log evidence and match them against deduplicated contracts. Public cross-chain spillovers are rare; the two alignable cases provide a \emph{lower-bound} check of whether the high-score queue hits real incidents.

\subsubsection{Percentile localization: real incidents hit the high-score queue}
Table~\ref{tab:event_alignment} reports aligned samples and their within-chain percentiles (cross-chain spillover cases). During screening, we identified two cross-chain incident samples that fall into the p99 tail but not the p99.9 extreme tail. The picture is consistent: p99 acts as the practical main audit queue, while p99.9 is best reserved for the most urgent, minimal queue. In practice, $\obfscore$ should be used for prioritization: review p99 tail first and combine with reuse clusters and low-signature evidence for deeper triage.

\begin{table*}[!t]
  \centering
  \caption{\textbf{Aligned cross-chain spillover cases with percentile localization. A single incident may appear multiple times with different evidence types.}}
  \label{tab:event_alignment}
  \scriptsize
  \setlength{\tabcolsep}{3pt}
  \renewcommand{\arraystretch}{0.85}
  \begin{tabular}{p{3.2cm}|c|c|c|c}
    \toprule
    \textbf{Incident} & \textbf{Chain} & \textbf{Evidence} & \textbf{Percentile} & \textbf{In tail (p99/p99.9)} \\
    \midrule
    Transit Swap DEX Hack & BSC & tx\_resolved & p99.74 & Yes / No \\
    New Free DAO Flash Loan & BSC & direct & p99.21 & Yes / No \\
    \bottomrule
  \end{tabular}
\end{table*}

\subsubsection{Case studies of cross-chain spillovers (excerpt)}
\textbf{Transit Swap DEX Hack (2022-10-02, BSC, tx\_resolved, p99.74):} public analyses and on-chain fund flows indicate cross-chain fund migration, i.e., a \emph{cross-chain deployment + cross-chain fund migration} spillover type. Its p99 hit shows the high-score queue captures real cross-chain spillover cases.\\
\textbf{New Free DAO Flash Loan (2022-09-08, BSC, direct, p99.21):} public analyses indicate part of the proceeds moved cross-chain for laundering, i.e., a \emph{single-chain exploit + cross-chain laundering} spillover type; it lands in p99 but not p99.9, reinforcing p99 as the practical main queue.

\medskip
\begin{center}
\fcolorbox{gray}{gray!10}{%
\parbox{0.95\linewidth}{\textbf{Finding 4 (RQ4).}
The aligned cross-chain spillover samples fall into the p99 tail but not p99.9, indicating that the high-score queue has \emph{real-world hit value} and that p99 is the practical main audit queue. Together with the small$\rightarrow$large diffusion in RQ3, this implies that high-obfuscation templates can spill over and lead to costly incidents. In practice, use p99 as the main queue, keep p99.9 for emergencies, and trigger cross-chain tracing once a high-obfuscation reuse cluster appears on smaller chains.}}
\end{center}

\subsection{Summary}
This chapter builds a coherent audit narrative across RQ1--RQ4. RQ1 establishes systematic cross-chain drift in obfuscation scores, showing that a single cutoff is not transferable and motivating within-chain percentile queues (p99 as the main queue and p99.9 as the emergency queue). RQ2 then translates the high-score queue into an interpretable structural profile—rare selectors, external-call enrichment, and low signature density—enabling secondary triage within the queue. RQ3 demonstrates directional template reuse across chains, closer to diffusion from smaller ecosystems to larger ones, making cross-chain joint auditing necessary. Finally, RQ4 closes the loop with incident alignment and spillover evidence, showing that the high-score queue has real-world hit value and can serve as a practical audit-prioritization signal. Overall, the chapter connects “distribution drift → structural profile → cross-chain propagation → real-world hits,” yielding actionable thresholds and a workflow for cross-chain auditing.

%% file: Text/related_work.tex
\section{Related Work}
\noindent\textbf{Obfuscation measurement and impact.}
Sheng et al.\ propose ObfProbe to quantify transfer-path obfuscation in Ethereum bytecode, defining a feature-based taxonomy and using a Z-score to rank obfuscation strength, and they show that heavy obfuscation concentrates risk and degrades detector effectiveness~\cite{sheng2025obfuscated}. Yang et al.\ further demonstrate that closed-source and control-flow obfuscation can conceal critical vulnerabilities in MEV bots, motivating deobfuscation-aware analysis~\cite{yang2025obscurity}. From an adversarial perspective, source-level obfuscation and bytecode-level obfuscation have been shown to substantially reduce the success of existing analyzers and decompilers~\cite{zhang2020obfuscation,yu2022bosc}. BiAn further systematizes source-level obfuscation and substantially increases decompilation and detection difficulty~\cite{zhang2023bian}. Together, these studies establish obfuscation as a security-relevant signal, yet they are largely Ethereum-centric and do not address whether obfuscation signals transfer across chains.

\noindent\textbf{Bytecode analysis and decompilation.}
Large-scale bytecode analysis is enabled by advances in decompilation and function recovery. Gigahorse introduced a declarative decompiler that lifts EVM bytecode to higher-level Intermediate Representation and supports scalable analyses~\cite{grech2019gigahorse}. Elipmoc improves recovery for complex control flow and optimized bytecode, while Shrnkr further increases scalability with shrinking context sensitivity~\cite{grech2022elipmoc,lagouvardos2025shrnkr}. Neural-FEBI improves function identification directly on EVM bytecode, enabling more precise downstream analyses~\cite{he2023neuralfebi,slither2019,securify2018,zeus2018,diangelo2024evolution}. These tools make obfuscation measurement feasible at scale and enable reuse analysis, but they are not designed to study cross-chain drift and have not been systematically applied to cross-chain distribution and reuse measurement.

\noindent\textbf{Learning-based program analysis and surrogates.}
Recent work applies machine learning to EVM bytecode for function identification and vulnerability detection~\cite{he2023neuralfebi,huang2021graph}. In parallel, transaction language models and graph-based representations have been used for Ethereum fraud detection and blockchain phishing detection~\cite{sun2025fraud,sheng2025dff}, highlighting the effectiveness of learned signals for security tasks. For vulnerability detection, SymGPT combines symbolic execution with LLMs~\cite{xia2025symgpt}, VASCOT applies Transformers to bytecode sequences~\cite{balci2025vascot}, and MANDO-HGT uses heterogeneous graph Transformers to capture control/data-flow semantics~\cite{liao2025mando,byteeye2026ase}. These approaches show that learned representations can reduce reliance on costly analysis pipelines, but they are not designed to score obfuscation or to support multi-chain distribution comparison. Our surrogate model follows this learning-based paradigm to enable fast obfuscation scoring at cross-chain scale.

\noindent\textbf{Code reuse and clone detection.}
Clone detection studies reveal pervasive code reuse in Ethereum, both at the contract level and the function level~\cite{he2020clones,khan2023cloning}. Chen et al.\ analyze code reuse patterns and show that standardized templates and shared components are prevalent in smart contract development~\cite{chen2021codereuse}. Recent work benchmarks clone detectors and compares their effectiveness on smart contracts~\cite{wang2025clonedetection}, and interpretable similarity tools have been proposed to identify reuse families~\cite{liu2025clone}. Clone propagation at the platform level has also been studied; BlockScope identifies cloned vulnerabilities across forked blockchains~\cite{yi2023blockscope}. These works imply that code reuse can propagate vulnerabilities and design patterns, yet they focus on single-chain ecosystems and do not quantify reuse families in cross-chain settings~\cite{duan2025pdlogger}.

\noindent\textbf{Cross-chain reuse and migration.}
In multi-chain contexts, EquivGuard reports that copy-and-paste reuse can introduce EVM-inequivalent code smells, highlighting semantic drift even among reused contracts~\cite{wang2025equivguard}. Empirical studies of sniper bots across Ethereum and BNB Smart Chain show that adversarial templates migrate between chains, suggesting cross-chain propagation of risky patterns~\cite{cernera2025sniperbots}. Cross-chain imitation attacks further demonstrate real-world threats from rapid cross-chain copying~\cite{qin2023imitation}. These findings motivate our focus on cross-chain reuse families in high-obfuscation tails and on the transferability of obfuscation thresholds.

\noindent\textbf{Cross-chain security context.}
Cross-chain interoperability and bridge systems are major risk concentrations~\cite{augusto2024sok,belenkov2025sok,chainalysis2022bridges}. Industry reports and tracing studies further show that cross-chain fund migration and laundering are widespread~\cite{trm2023illicit,lin2025abctracer,yan2025ledger,crosschain2020}. Public enforcement actions document multi-chain scam deployments, such as the Forsage case~\cite{sec2022forsage}. These risks compound with cross-chain reuse, which raises the stakes for auditing and triage. These works provide incident and ecosystem context, but they do not analyze contract-level obfuscation; our work aligns high-obfuscation signals with such incidents to validate practical relevance.

\begin{center}
\fcolorbox{gray}{gray!10}{%
\parbox{0.95\linewidth}{\textbf{Positioning and comparison.}
Compared with ObfProbe and other Ethereum-centric obfuscation studies~\cite{sheng2025obfuscated,yang2025obscurity}, we focus on cross-chain transferability by quantifying distribution shifts and tail composition across multiple EVM chains rather than a single-chain prevalence analysis. Relative to bytecode analysis and decompilation work~\cite{grech2019gigahorse,grech2022elipmoc,lagouvardos2025shrnkr,he2023neuralfebi}, our contribution is not a new lifter but a scalable measurement pipeline that uses learned scoring to accelerate cross-chain analysis. In contrast to clone and reuse studies that remain largely single-chain~\cite{he2020clones,khan2023cloning,chen2021codereuse,wang2025clonedetection,liu2025clone}, we explicitly trace reuse families across chains and examine whether high-obfuscation tails are enriched by cross-chain reuse. Finally, while bridge security studies and incident reports quantify cross-chain risk~\cite{augusto2024sok,belenkov2025sok,sec2022forsage}, our work links those risks to obfuscation signals, offering a measurement-based view of where cross-chain auditing should prioritize attention.}}
\end{center}

%% file: Text/discussion.tex
\section{Discussion}

\subsection{Operational Implications and Audit Workflow}
Our results transfer cross-chain obfuscation measurements into an \emph{actionable audit workflow}. RQ1 provides a two-tier queueing strategy: p99 as the main queue for stable coverage, and p99.9 as an emergency queue for urgent triage. RQ2’s structural profile (rare selectors, external-call enrichment, low signature density) enables secondary prioritization within the main queue, focusing scarce analyst effort on contracts with the highest interpretation cost. RQ3 motivates cross-chain joint auditing: once a high-score reuse cluster is found on one chain, the same bytecode hash should be searched on other chains. RQ4’s incident alignment and case studies show that the p99 queue has real-world hit value and can be used as a practical \emph{prioritization signal}.
\begin{center}
\fcolorbox{gray}{gray!10}{%
\parbox{0.95\linewidth}{
\textbf{Minimal SOP (Audit Workflow)} \\
Step 1 (Queueing): use within-chain p99 for the main queue and p99.9 for the emergency queue.\\
Step 2 (Secondary triage): re-rank the p99 queue by “low signature density + rare selectors + external-call enrichment.”\\
Step 3 (Cross-chain linkage): for high-score reuse clusters (same bytecode hash), search and aggregate hits on other chains.\\
Step 4 (Evidence closure): if incident signals or tx/log evidence is found, escalate to manual audit and reporting.
}}
\end{center}

\subsection{Mechanistic Interpretation of Drift and Spillover}
Score drift and tail-structure differences are driven by ecosystem-specific contract composition and template reuse. The enrichment of low-visibility logic and complex external-call patterns in the tail suggests that high-score regions are dominated by template-like, hard-to-interpret contracts. Directional reuse indicates that high-score templates diffuse across chains, more often from smaller ecosystems to larger ones, which aligns with a spillover risk channel. This mechanism explains why a single global threshold fails and why cross-chain coordination is needed.
\textbf{H1 (Template Diffusion Hypothesis):} high-obfuscation templates are more likely to be reused across chains and to diffuse toward larger ecosystems (evidence: tail Jaccard enrichment and directional overlap).\\
\textbf{H2 (Low-Visibility Structure Hypothesis):} the tail is driven by low-visibility logic and complex call chains (evidence: selector/opcode enrichment and low signature density).

\subsection{Scope and Risk Boundaries}
\obfscore is a screening and ranking signal rather than a direct “maliciousness” label. The extreme threshold (p99.9) is best viewed as an emergency queue; relying solely on it can miss real incidents. Incident alignment provides evidence of real-world hit value but remains constrained by incomplete public reports and evidence-chain availability. Thus, our contribution is a \emph{prioritization workflow}, not a one-shot classifier.
In addition, cross-chain comparison should not directly reuse absolute scores from another chain; and scores should be combined with reuse, visibility, and call-structure cues, otherwise alert fatigue and audit-resource contention can occur.

\subsection{Limitations and Future Work}
The surrogate inherits noise from \textsc{Obfs\_Tool} labels and does not directly capture semantic risk. Cross-chain calibration can be further refined. Future work can proceed in two concrete directions: (i) cross-chain calibration via quantile mapping or monotonic calibration functions that map $\obfscore$ into unified risk bands; (ii) semantic enhancement by integrating verified source code, known exploit labels, and audit reports into a multi-task “obfuscation + risk semantics” model. More fine-grained template families and interaction graphs are also promising to improve interpretability and cross-chain early warning.

%% file: Text/conclusion.tex
\section{Conclusion}
We present \textsc{HObfNET}, a fast surrogate of \textsc{Obfs\_Tool} whose inference cost and speed support million-scale, cross-chain scoring. We analyze four chains at scale (ETH 1,042,923; BSC 2,308,899; Polygon 288,611; Avalanche 96,173) and provide operational, \emph{within-chain} thresholds (p99/p99.9; e.g., ETH 18.07/22.69, BSC 16.82/19.74, Polygon 18.72/20.51, Avalanche 19.18/20.67). On top of Ethereum supervision, we answer four research questions: (i) scores drift across chains; (ii) the high-score tail exhibits a structural profile with rare selectors, external-call enrichment, and low signature density, supporting secondary triage; (iii) high-score templates show directional cross-chain reuse, with tail Jaccard about 1.5--2.0$\times$ higher than overall; and (iv) incident alignment on the available cross-chain cases (all in p99, none in p99.9) shows real-world hit value of the high-score queue within the aligned evidence set. Collectively, these results yield an actionable cross-chain audit workflow and prioritization queues for security operations.

Future work includes cross-chain calibration that maps $\obfscore$ into unified risk bands and semantic enhancement by integrating verified source code/verified contracts, known exploit labels, and audit reports to improve interpretability and early warning.